\documentclass[prd,groupedaddress]{revtex4}
\usepackage{graphicx}
\usepackage{graphics}
\usepackage{amsmath}
\usepackage{amssymb}
\usepackage{mathtools} 

\begin{document}

\title{Renormalizability in $D$-dimensional higher-order gravity } 

\author{Antonio Accioly}

\email[]{accioly@cbpf.br}

\affiliation{Coordena\c{c}\~{a}o de Cosmologia, Astrof\'{i}sica e Intera\c{c}\~{o}es Fundamentais (COSMO), \\ Centro Brasileiro de Pesquisas F\'{i}sicas (CBPF), Rua Dr. Xavier Sigaud 150, Urca, 22290-180, \\ Rio de Janeiro, RJ, Brazil}

\author{Jos\'{e} de Almeida}

\email[]{josejr@cbpf.br}

\affiliation{Coordena\c{c}\~{a}o de Cosmologia, Astrof\'{i}sica e Intera\c{c}\~{o}es Fundamentais (COSMO), \\ Centro Brasileiro de Pesquisas F\'{i}sicas (CBPF), Rua Dr. Xavier Sigaud 150, Urca, 22290-180, \\ Rio de Janeiro, RJ, Brazil}

\author{Gustavo P. Brito}

\email[]{gpbrito@cbpf.br}

\affiliation{Coordena\c{c}\~{a}o de Cosmologia, Astrof\'{i}sica e Intera\c{c}\~{o}es Fundamentais (COSMO), \\ Centro Brasileiro de Pesquisas F\'{i}sicas (CBPF), Rua Dr. Xavier Sigaud 150, Urca, 22290-180, \\ Rio de Janeiro, RJ, Brazil}

\author{Gilson Correia}

\email[]{gilson@cbpf.br}

\affiliation{Coordena\c{c}\~{a}o de Cosmologia, Astrof\'{i}sica e Intera\c{c}\~{o}es Fundamentais (COSMO), \\ Centro Brasileiro de Pesquisas F\'{i}sicas (CBPF), Rua Dr. Xavier Sigaud 150, Urca, 22290-180, \\ Rio de Janeiro, RJ, Brazil}

\date{\today}


\begin{abstract} 
A simple expression for calculating the classical potential concerning $D$-dimensional gravitational models is obtained through a method based on the generating functional. The prescription is then used  as a mathematical tool to probe  the conjecture that renormalizable higher-order gravity models --- which are, of course, nonunitary --- are endowed with  a classical potential that is nonsingular at the  origin. It is also shown that the converse of this statement is not true, which implies that the finiteness of the classical potential at the origin is a necessary but not a sufficient condition for the renormalizability of the model. The systems we have  utilized to verify the conjecture were fourth- and sixth- order gravity models in $D$-dimensions.  A discussion about the polemic question related to the renormalizability of new massive gravity, which Oda claimed to be renormalizable in 2009 and three years late was shown to be nonrenormalizable by Muneyuki and Ohta, is considered. We remark that the solution of this issue is straightforward if the aforementioned conjecture is employed. We point out that
 our analysis is restricted to local models in which the propagator has simple  and real poles.

\end{abstract}


\maketitle

\section{Introduction}

Higher-order gravity models are prime candidates as far as the construction of a renormalizable gravity theory is concerned. In fact, the higher-order terms of these systems are responsible in general for taming the wild ultraviolet divergences present in the Einstein-Hilbert action. In addition, as is well known, a pacific coexistence between renormalizability and unitarity is generally unattained in these models.

Recently, many authors \cite{1,2,3,4,5,6,7,8,9,10,11,12,13,14,15,16,17,18,19} have  addressed themselves to the problem of verifying a conjecture that --- as far as we know ---was hinted for the first time by Stelle \cite{20,21} in his analysis of the renormalizability of fourth-order gravity in four dimensions:  Renormalizable higher-order gravity models, are endowed with a classical potential lacking a singularity at the origin. Nonetheless, neither Stelle  nor the subsequent authors up to now   seemed to perceive in their guesstimates  that the converse of this premise is not true.

Our main goal here is exactly to probe via some specific models that  the finiteness  of the potential at the origin is a necessary but not a sufficient condition for the renormalizabity of the model. 

A natural question must then be posed. What is the utility of this conjecture? 
 The advantages that result from this surmise are very relevant. Indeed, by simply computing the classical potential at the origin we can be absolutely certain that any higher-derivative gravity model with a divergent potential at the origin is nonrenormalizable. An more, if we are uncertain about the renormalizability of a given system as is the case of New Massive Gravity  (NMG) \cite{22,23,24,25}, which Oda \cite{26} claimed to be renormalizable and three years later Muneyuki and Ohta \cite{27} showed to be nonrenormalizable, using our conjecture we would promptly conclude that this system  is nonrenormalizable since its gravitational  potential is singular at the origin. If we make a detailed comparison between the simplicity of our premise and the difficult computations required by the ordinary methods of  Quantum Field Theory, we come to the conclusion that our surmise is much easier to handle in the cases just mentioned. It is important to recall that the task of proving the renormalizability of a given higher-order gravity model is a very hard work even for the experts on the subject, which can be easily seen by leafing through the aforementioned articles \cite{26,27}, as well as the ones by Stelle \cite{20}, Antoniadis and Tonboulis \cite{28}, and Johnston \cite{29}.

The models we shall use to probe the mentioned conjecture are fourth- and sixth- order gravity systems in $D$-dimensions, and  a   particular sixth-order  gravity system in four dimensions. They are defined  by the following actions

\begin{eqnarray}
I^{({\mathrm{fourth-order}})}= \int d^D x \sqrt{|g|} \Bigg[ \frac{2\sigma}{\kappa^2}R + \frac{\alpha }{2}R^2 + \frac{\beta}{2}R^2_{\mu \nu} \nonumber + \frac{\gamma}{2} R^2_{\mu \nu \alpha \beta}   - {\cal L}_{\mathrm{M}} \Bigg],
\end{eqnarray}

\begin{eqnarray}
I^{({\mathrm{sixth-order}})}\!\!=\!\! \int d^D x\! \sqrt{|g|} \frac{1}{\kappa^2} \Bigg[ 2R \!+\! \frac{\alpha_0 }{2}R^2 \!+\! \frac{\beta_0}{2}R^2_{\mu \nu} \!+\!  
 \frac{\gamma_0}{2} R^2_{\mu \nu \alpha \beta}   \!+\! \frac{\alpha_1}{2} R \Box R  \!+\! \frac{\beta_1}{2} R_{\mu \nu} \Box R^{\mu \nu }  \!+\! \frac{\gamma_1}{2} R_{\mu \nu \alpha \beta} \Box R^{\mu \nu \alpha \beta} \!-\! {\cal L}_{\mathrm{M}} \Bigg],
\end{eqnarray}

\begin{eqnarray}
I= \int d^4 x\sqrt{|g|} \Big( \frac{2}{\kappa^2} R + {\alpha'_0} R^2  + a'_1 R \Box R + b'_0 R^2_{\mu \nu}
 - {\cal L}_{\mathrm{M}} \Big), \nonumber
\end{eqnarray}

\noindent where $\sigma = \pm 1, \alpha, \beta, \gamma, \alpha_0, \beta_0, \gamma_0, \alpha_1, \beta_1, \gamma_1,  \alpha'_0, a'_1,  b'_0$ are arbitrary constants,  $\kappa^2=4\kappa_D$, and ${\cal L}_{\mathrm{M}}$ is the Lagrangian for matter, being

\begin{eqnarray}
\kappa_D= \Bigg( \frac{D-2}{D-3} \Bigg) G_D \frac{2\pi^{\frac{D-1}{2}}}{\Gamma\Big(\frac{D-1}{2}\Big)},
\end{eqnarray}
\noindent the $D$-dimensional Einstein constant for $D >3$ (see Appendix A). Here $G_D$ is the Newton constant in $D$-dimensions ($D >3$), and $\Gamma$ is the gamma function. Note that $\kappa_D$ reduces to its usual value in four dimensions, namely $\kappa_D=8 \pi G_4$. We remark also that the Einstein constant in $D=3$ cannot be related to $G_3$ since general relativity in three dimensions is trivial and, as a consequence, has no Newtonian limit. Nevertheless, for simplicity's sake $\kappa_3$ will be used from now on as the symbol for the Einstein constant in $D=3$, although it is unrelated to $G_3$.

Now, since in order to probe the conjecture at hand we are required to compute the gravitational potential, the efficiency with which we will make the verification of this surmise will heavily depend on how skilled we are in building out a simple prescription for calculating this potential.   Accordingly, in Sec. II we construct a straightforward method for calculating the $D$-dimensional   gravity potential based on the generating functional . Using this prescription the conjecture is verified for fourth- and sixth- order gravity models in $D$-dimensions in Secs. III and IV, respectively. We point that the analysis of  the tree-level unitary  of the aforementioned systems is made in the respective sections. The aim of this study is to confirm the general premise that renormalizable higher-order models are nonunitary. We present our conclusions in Sec. V. We remark also that in this last section  a special attention is devoted to NMG since it was the analysis of this model that inspired our conjecture.

It worth mentioning that we will only deal with local models in which the poles are simple and real.

Technical details will be relegated to the Appendices.

We use natural units throughout and our Minkowski  metric is diag(1, -1, -1, ..., -1).

\section{Simple prescription for calculating the $D$-dimensional potential for gravitational models}

From Quantum Field Theory we know that  the generating functional for the  connected Feynman diagrams $W_D(T)$ is related to the generating functional $Z_D(T)$ for linearized gravity theories   by   $Z_D(T)= e^{i W_D(T)}$ \cite{30,31,32},
\noindent where 

\begin{eqnarray}
W_D(T)= -\frac{\kappa_D}{2} \int d^D x d^D y T^{\mu \nu}(x) D_{\mu \nu, \alpha \beta }(x-y) \times T^{\alpha \beta}(y).
\end{eqnarray}  

\noindent Here $T^{\mu \nu }(x) \;(=T^{\nu \mu}(x))$ and $D_{\mu \nu, \alpha \beta}(x-y)$ are, respectively, the external conserved  current and the propagator.

Now, keeping in mind that 

\begin{eqnarray}
D_{\mu \nu, \alpha \beta}(x-y)&=& \int{\frac{d^D k}{(2\pi)^D}e^{ik(x-y)}D_{\mu \nu, \alpha \beta}(k)},\nonumber \\ T^{\mu \nu}(k)&=& \int{d^D x e^{-ikx}}T^{\mu \nu}(x), \nonumber
\end{eqnarray}

\noindent we promptly obtain

\begin{eqnarray}
W_D(T)= -\frac{\kappa_D}{2} \int{\frac{d^D k}{(2\pi)^D} T^{\mu \nu}(k)^*{\cal P}_{\mu \nu, \alpha \beta}(k)T^{\alpha \beta}(k)}, \nonumber
\end{eqnarray}

\noindent where ${\cal P}_{\mu \nu, \alpha \beta}(k)$ is the `modified propagator' in momentum space obtained by neglecting all terms of the usual Feynman propagator that are orthogonal to the external conserved currents

Assuming then that the external  conserved current is time independent, we get from the preceding equation

\begin{eqnarray}
W_D(J)=- \frac{\kappa_D}{2} \int \frac{d^D k}{(2 \pi)^{D-1}} \Big[ \delta (k^0) \; T \;  {\cal P}_{\mu \nu, \alpha \beta}(k)  \int \int  d^{D-1} {\bf{x}}   
d^{D-1} {\bf{y}} e^{i{\bf{k}}\cdot{\bf{(y-x})}}T^{\mu \nu}({ \bf x)}T^{\alpha \beta}({\bf y}) \Big],
\end{eqnarray}

\noindent where the time interval $T$ is produced by the factor $\int{dx^0}$.

Simple algebraic  manipulations, on the other hand, reduces (5) to the form

\begin{eqnarray}
W_D(T)= -\kappa_DT\int{\frac{d^{D-1} {\bf{k}}}{(2 \pi)^{D-1}} {\cal P}_{\mu \nu, \alpha \beta}({\bf k}) \Delta^{\mu \nu, \alpha \beta}({\bf{k}})},
\end{eqnarray}

\noindent where  $ {\cal P}_{\mu \nu, \alpha \beta}({\bf k}) \equiv  {\cal P}_{\mu \nu, \alpha \beta}(k)|_{k^0=0} $,  and

\begin{eqnarray}
\Delta^{\mu \nu, \alpha \beta}({\bf{k}}) \equiv \int \int d^{D-1} {\bf{x}} d^{D-1} {\bf{y}} e^{i {\bf{k}} \cdot ({\bf{y-x})}} \frac{T^{\mu \nu}( {\bf x}) T^{\alpha \beta}({ \bf y})}{2}. \nonumber
\end{eqnarray}

 In the specific case of two  masses $M_1$ and $M_2$ located, respectively, at ${\bf{a_1}}$ and ${\bf{a_2}}$, the current assumes the form

\begin{eqnarray}
T^{\mu \nu}({\bf x}) = \eta^{\mu 0} \eta^{\nu 0}\Big[ M_1\delta^{D-1}({\bf{x - a_1}})
 + M_2 \delta^{D-1} ({\bf{x - a_2}})\Big]. \nonumber 
\end{eqnarray}

Therefore, 

\begin{eqnarray}
\Delta^{\mu \nu, \alpha \beta}({\bf{k}})= M_1 M_2 e^{i {\bf{k}} \cdot {\bf{r}}} \eta^{\mu 0} \eta^{\nu 0} \eta^{\alpha 0} \eta^{\beta 0},
\end{eqnarray}

\noindent where ${\bf{r= a_2 - a_1}}$. 

As a consequence,

\begin{eqnarray}
W_D(T)= - \kappa_D T\frac{M_1 M_2}{(2 \pi)^{D-1}} \int{d^{D-1}{\bf{k}}e^{i {\bf{k}} \cdot {\bf{r}}} {\cal P}_{00,00}({\bf{k}})}.
\end{eqnarray}

Bearing in mind that

\begin{eqnarray}
Z_D(T)= <0\big|e^{-iH_D T} \big|0> = e^{-iE_DT},
\end{eqnarray}

\noindent which implies that 

\begin{eqnarray}
E_D = - \frac{W_D(T)}{T},
\end{eqnarray}

\noindent we find that  the $D$-dimensional   interparticle gravitational energy can be computed through the simple expression

\begin{eqnarray}
E_D(r)=  \kappa_D\frac{M_1 M_2}{(2 \pi)^{D-1}} \int{d^{D-1}{\bf{k}}e^{i {\bf{k}} \cdot {\bf{r}}}
 {\cal P}_{00,00}({\bf{k}})}.
\end{eqnarray}

Accordingly, the $D$-dimensional gravitational potential sourced by a mass $M$ at rest is given by

\begin{eqnarray}
V_D(r)= \kappa_D \frac{M}{(2\pi)^{D-1}} \int d^{D-1}{\bf k}{}e^{i {\bf{k}} \cdot {\bf{r}}}{\cal P}_{00, 00}({\bf k}).
\end{eqnarray}

Using the straightforward prescription above it is possible to test the aforementioned conjecture easily, as will be shown in the next two sections.

\section{Verifying the conjecture for Fourth-order gravity systems in $D$-dimensions}
To find the gravitational potential we need beforehand to compute the propagator. Nonetheless, before obtaining this operator it is worthwhile remembering that this calculation demands only the knowledge of the linearized quadratic part of the model. On the other hand, since linearized Gauss-Bonnet invariant is a total derivative in any spacetime dimension $> 3$ (the restriction to $D=4$ coming in only when we take the full nonlinear structure into account) \cite{33}, and in addition both the curvature and Ricci tensors have the same number of components in $D=3$ \cite{34},  we can drop  the term of action (1) containing  $R^2_{\mu \nu \alpha \beta}$  for $D> 2$ in the mentioned computation.

To compute the propagator we recall that for small fluctuations around the Minkowski metric $\eta_{\mu \nu}$, the full metric assumes the form

\begin{eqnarray}
g_{\mu \nu}= \eta_{\mu \nu} + \kappa h_{\mu \nu}.
\end{eqnarray}

Linearizing the Lagrangian associated with the quadratic part of the action (1), namely

\begin{eqnarray}
{\cal{L}}^{({\mathrm{fourth-order}})}= \sqrt{|g|} \Bigg[ \frac{2\sigma}{\kappa^2}R + \frac{\alpha }{2}R^2 + \frac{\beta}{2}R^2_{\mu \nu}  \Bigg],
\end{eqnarray}

\noindent via the preceding equation and adding to the result the gauge-fixing Lagrangian, ${\cal L}_{\mathrm{gf}}= \frac{1}{2\lambda}(\partial_\mu \gamma^{\mu \nu})^2$, where $\gamma_{\mu \nu} \equiv h_{\mu \nu} - \frac{1}{2} \eta_{\mu \nu}h$ and $\lambda$ is a gauge parameter (de Donder gauge), we find

\begin{eqnarray}
{\cal{L}}^{({\mathrm{fourth-order}})}= \frac{1}{2}h_{\mu \nu}{\cal O}^{\mu \nu, \alpha \beta}h_{\alpha \beta},
\end{eqnarray}
\noindent where, in momentum space, 

\begin{eqnarray}
{\cal{O}}=&&  \Bigg(\sigma + \frac{\beta \kappa^2 k^2}{4}\Bigg)k^2 P^{(2)} + \frac{k^2}{2 \lambda} P^{(1)} + \frac{k^2}{4 \lambda}P^{(0-w)} - \frac{k^2}{4\lambda}\sqrt{D-1}\Bigg[P^{(0-sw)} + P^{(0-ws)} \Bigg] \nonumber \\  +&& \Bigg[ - (D-2)\sigma + (D-1) \alpha \kappa^2 k^2 + D \frac{\beta \kappa^2 k^2}{4} +  \frac{D-1}{4\lambda}\Bigg]k^2 P^{(0-s)}.
\end{eqnarray}

Inverting this operator we obtain the propagator for fourth-order gravity in $D$-dimensions, i.e.

\begin{eqnarray}
D^{(\mathrm{fourth-order})} = && \frac{1}{\sigma}\Bigg[ \frac{1}{k^2} - \frac{1}{k^2 - m_2^2}\Bigg] P^{(2)} +\frac{2\lambda}{k^2}P^{(1)} +  \frac{1}{\sigma (D-2)}\Bigg[    \frac{1}{k^2- m_0^2 }   -\frac{1}{k^2}  \Bigg ] P^{(0-s)} \nonumber \\  && + \Bigg[ \frac{4 \lambda}{k^2} + \frac{(D-1)m_0^2}{  \sigma k^2( k^2- m_0^2)(D-2)} \Bigg] P^{(0-w)} + \frac{\sqrt{D-1} m_0^2}{(D-2) \sigma  k^2 (k^2 - m_0^2)} \Bigg[ P^{(0-sw)}  + P^{(0-ws)} \Bigg],
\end{eqnarray}

\noindent where $\{ P^{(1)}, P^{(2)}, ..., P^{(0-ws)}\}$ is the set of the usual $D$-dimensional Barnes-Rivers operators (see Appendix B), and

\begin{eqnarray}
m^2_2\equiv - \frac{4 \sigma}{\beta \kappa^2}, m^2_0 \equiv \frac{4\sigma(D-2)}{\kappa^2 \Big[4 \alpha (D-1) + D \beta\Big]}.
\end{eqnarray}

\noindent Here we are supposing that there are no tachyons in the model, which implies that $m^2_2 >0$  and $m^2_0 >0$.

The expression for the spatial part of the modified propagator can be trivially found by means of (17). Making the  appropriate computations we arrive at the following result

\begin{eqnarray}
{\cal P}_{\mu \nu, \alpha \beta}({\bf k})\!=\! \frac{1}{\sigma}\Bigg\{ \Bigg[\!- \frac{1}{{\bf k}^2}
\!+\! \frac{1}{{\bf k}^2 + m^2_2}\Bigg] \Big[ \frac{1}{2}(\eta_{\mu \kappa} \eta_{\nu \lambda} \!+\! \eta_{\mu \lambda} \eta_{\nu \kappa})  \!-\! \frac{1}{D-1} \eta_{\mu \nu} \eta_{\kappa \lambda} \Big] \!+\! \frac{\eta_{\mu \nu} \eta_{\kappa \lambda}}{(D-1)(D-2)} \Bigg[ \frac{1}{{\bf k}^2} \!-\! \frac{1}{{\bf k}^2 + m^2_0}\Bigg] \Bigg \}.
\end{eqnarray}

As a consequence,

\begin{eqnarray}
{\cal P}_{00, 00}({\bf k})= \frac{1}{\sigma} \Bigg( - \frac{D-3}{D-2} \frac{1}{{\bf k}^2}
+ \frac{D-2}{D-1} \frac{1}{{\bf k}^2 + m^2_2}-  \frac{1}{(D-1)(D-2)({\bf k}^2 + m^2_0)} \Bigg).
\end{eqnarray}

Therefore, the $D$-dimensional gravitational potential generated by a static mass $M$ can be computed through the expression
\begin{align}
V^{( \mathrm{fourth-order})}_D (r)&= - \frac{\kappa_D M}{\sigma (2 \pi)^{D-1}} \Bigg[ \frac{D-3}{D-2} \int \frac{d^{D-1} {\bf k}}{{\bf k}^2} e^{i {\bf k} \cdot {\bf r}} + \\&- \frac{D-2}{D-1} \int \frac{d^{D-1} {\bf k}}{{\bf k}^2 + m^2_2} e^{i {\bf k} \cdot {\bf r}} + \frac{1}{(D-2)(D-1 )} \int \frac{d^{D-1} {\bf k}}{{\bf k}^2 + m^2_0} e^{i {\bf k} \cdot {\bf r}}  \Bigg]. 
\end{align}

Performing the integrations, we find (see Appendix C)

\begin{eqnarray}
V^{( \mathrm{fourth-order})}_D (r)=&& - \frac{\kappa_D M}{\sigma (2 \pi)^{\frac{D-1}{2}} }\Bigg[ \frac{D-3}{D-2} \frac{2^{\frac{D-5}{2}}}{r^{D-3}} \Gamma \Big( \frac{D-3}{2} \Big)  - \frac{D-2}{D-1}\Big( \frac{m_2}{r} \Big)^{\frac{D-3}{2}} K_{\frac{D-3}{2}} (m_2 r) \nonumber \\ && + \frac{1}{(D-1)(D-2)} \Big(\frac{m_0}{r} \Big)^{\frac{D-3}{2}} K_{\frac{D-3}{2}}(m_0 r) \Bigg],  \;\; (D=4,5)
\end{eqnarray}
 
\noindent and 

\begin{eqnarray}
V^{({ \mathrm {fourth-order}})}_3(r)= \frac{\kappa_3 M}{4 \pi \sigma} \Big[K_0 (m_2 r)- 
K_0(m_0 r) \Big],
\end{eqnarray}
 \noindent  wherein $K_\nu$ is the modified Bessel function of the second order of order $\nu$.

Bearing in mind that

\begin{eqnarray}
K_\nu (r) \sim \sqrt{\frac{\pi}{2}} \frac{e^{-r}}{\sqrt{r}} \Bigg(1 + {\cal O}\Big(\frac{1}{r}\Big) \Bigg) \;\ \; (r \rightarrow \infty),
\end{eqnarray}

\noindent it is trivial to see that (21) and the Newton gravitational potential agree asymptotically if and only if $\sigma = +1$. Accordingly,  we assume from now on that $\sigma = +1$ for $D>3$.

Before going on, it is important to call attention to the fact that our discussion will be restricted to the systems in 3, 4, and 5 dimensions since these are the only models in which
 is possible to compute the gravitational potential analytically.

We analyze now the small-distance behavior of the gravitational potential concerning the mentioned systems.

\subsection{$D=3$}

Remembering that for $x \ll 1$,

\begin{eqnarray}
K_0(x) \sim- \Bigg( \gamma + \ln\frac{x}{2}\Bigg) + \frac{x^2}{4}\Bigg(1- \gamma - \ln \frac{x}{2} \Bigg) + x^4 \Bigg( \frac{1}{128}(3- 2\gamma) - \frac{1}{64} \ln \frac{x}{2} \Bigg) + ...,
\end{eqnarray}

\noindent where $\gamma$ is the Euler-Mascheroni constant, we may rewrite the expression for the gravitational potential (22) as

\begin{eqnarray}
V_3(r)\sim \frac{\kappa_3 M}{4 \pi \sigma} \Bigg[ \ln \frac{m_0}{m_2} + \frac{(m_2 r)^2}{4 } \Bigg( 1 - \gamma \ln \frac{m_2 r}{2} \Bigg) - \frac{(m_0 r)^2}{4}\Bigg( 1 - \gamma \ln \frac{m_0 r}{2} \Bigg)  + ... \Bigg].
\end{eqnarray}

Thence, as $r \rightarrow 0$, we get

\begin{eqnarray}
V_3(0)= \frac{\kappa_3 M}{4 \pi \sigma }\ln \frac{m_0}{m_2}.
\end{eqnarray}

It follows then that full tridimensional fourth-order gravity theories, i.e. the models with no special relations between their parameters, have a gravitational potential that is finite at the origin. However, NMG \cite{22}, for instance, where their parameters are linked via the constraint $8\alpha + 3 \beta$, is singular at the origin. Note that $\sigma$ for this system is equal to  -1. We shall analyze the alluded model in Sec. V.

\subsection{$D=4$} 

Taking into account that $K_{\frac{1}{2}}(x) = \sqrt{\frac{\pi}{2}} \frac{e^{-x}}{\sqrt{x}}$, we immediately obtain from (21)

\begin{eqnarray}
V_4(r)= - \frac{\kappa_4 M}{8 \pi r} \Bigg(1 - \frac{4}{3} e^{-m_2 r} + \frac{1}{3}e^{- m_0 r } \Bigg).
\end{eqnarray}

To check whether $V_4(r)$ is regular at the origin, we expand the exponentials at $r=0$ into power series. Doing so it is easy to verify that the contribution of the higher-derivative terms cancel the Newtonian one  making  the model  free of singularity. In fact, the alluded potential can be written as

\begin{eqnarray}
V_4(r)\sim MG_4 \frac{m_0 - 4m_2}{3} + {\cal O}(r).
\end{eqnarray}

\noindent The singularity cancellation occurs because the zero order terms containing higher-derivatives produce a coefficient +1 responsible for canceling out the coefficient  -1 from the original Newton term.

\subsection{D=5}
Keeping in mind that for $x \rightarrow 0$, 

\begin{eqnarray}
K_1(x) \sim \frac{1}{x} + \frac{x}{4}\Bigg[ 2 \gamma - 1 + \frac{1}{8} \Big(2 \gamma - \frac{5}{2}\Big) x^2 + \frac{1}{192} \Big(2 \gamma  \nonumber  - \frac{10}{3} \Big) x^4 + ... \Bigg] + \frac{x}{2} \ln\frac{x}{2} \Bigg[ 1 + \frac{x^2}{8} + \frac{x^4}{192} + ... \Bigg], \nonumber
\end{eqnarray}
 \noindent we may write $V_5(r)$ as 

\begin{eqnarray}
V_5(r)\sim - \frac{\kappa_5 M}{ 48 (2 \pi)^2 }\Bigg[ \Big( m^2_0 - 9m^2_2 \Big)\Big( 2 \gamma - 1  +2\ln r \Big) + m^2_0  \times \ln\frac{m^2_0}{4 }- 9m^2_2 \ln\frac{m^2_2}{4}  + ... \Bigg].
\end{eqnarray}

\noindent As a consequence, the full fourth-order gravitational potential in five dimensions is divergent at the origin; nevertheless, if $m^2_0 = 9m^2_2$, this potential is finite at the cited point. Accordingly, we have found  a nonsingular potential at the origin in five dimensions related to fourth-order gravity, being its value given by

\begin{eqnarray}
V_5(0)\Big|_{m^2_0= 9m^2_2} = - \frac{3\kappa_5 M m^2_2 \ln 3}{32 \pi^2}.
\end{eqnarray}

Let us then probe our conjecture for fourth-order gravity in $D$-dimensions.

\subsection{Testing the conjecture}
According to our conjecture the necessary condition for a $D$-dimensional higher-order model to be renormalizable is that it has a classical potential finite at the origin. As we have just shown, full fourth-order gravity systems in $D=3 ,4 $
 are finite at the origin, while in $D=5$ the full model has a singularity at the aforementioned  point. So, if the conjecture at hand is correct, both the three-  and four- dimensional full models are expected to be renormalizable, whereas the five-dimensional one should be nonrenormalizable. 

Now, since full fourth-order gravity models in $D= 3 , 4$ are known to be renormalizable \cite{20, 27} they agree with our conjecture since as have  just demonstrated,  they lack a singularity at the origin.

As far  as the  five-dimensional system is concerned,  it is trivial to show by power counting that the full model is nonrenormalizable. In fact, in this case the degree of superficial divergence is given by

\begin{eqnarray}
\delta= 5 + \frac{1}{2}\Bigg(\sum_{n=3}^{\infty}(n-2)(V_n - E)\Bigg),
\end{eqnarray}

\noindent which clearly shows that the system is nonrenormalizable since $\delta$ becomes greater as the  vertices number increases. Remembering that this model is divergent at the origin, it is in agreement with our  our surmise because it asserts that renormalizable systems must always be finite at the origin.

On the other hand, the gravitational potential concerning NMG is divergent at the origin, as we shall prove in Sec. V, while the five-dimensional model with its parameters 
connected by the relation $m^2_0= 9m^2_2$, has a potential that is free of singularity at the origin. Both systems are in accord with our conjecture. Indeed, new massive gravity is nonrenormalizable \cite{27} and the five-dimensional model is nonrenormalizable by power counting. 
 Note that our surmise says that the existence of a classical potential lacking a singularity at the origin is a necessary but not a sufficient condition for the renormalizability of the theory.

For completeness' sake, we discuss now the tree-level unitarity of the fourth-order gravity models.

\subsection{Unitarity of the fourth-order gravity systems} 
We show now that full fourth-order gravity models are nonunitaty in $D=3, 4, 5$. To do that we make use of a method pioneered by Veltman \cite{35} which has been extensively used since it was conceived. The prescription consists in saturating the propagator with conserved external currents and computing afterward the residues at the simple poles of the alluded saturated propagator ($SP$). If the residues at all  poles are positive or null, the system is tree-level unitary, but if at least one of the   residues is negative, the model is nonunitary at the tree level.

For $D=4$ and  $D=5$ we obtain from (19) the saturated propagator in momentum space (Note that we have chosen $\sigma= +1$ for the reasons already explained)

\begin{eqnarray} 
SP(k)= T_{\mu \nu}(k)D^{\mu \nu, \alpha \beta}(k)T_{\alpha \beta}(k) = \frac{A}{k^2} - \frac{B}{k^2 - m^2_2} + \frac{C}{k^2 - m^2_0}.  \nonumber
\end{eqnarray}

\noindent Here $$A\equiv T^2_{\mu \nu} - \frac{T^2}{2}, \; B \equiv T^2_{\mu \nu} - \frac{T^2}{3}, \; C \equiv \frac{T^2}{6},$$

 \noindent where $T_{\mu \nu}$ is an external conserved current, being $T_{\mu \nu}=T_{\nu \mu}$.

Now, taking into account that (see Ref. \cite{33} ) 

\begin{eqnarray}
\Big(T^2_{\mu \nu} - \frac{T^2}{2}\Big)\Big \vert_{k^2=0} >0&&, \;\;  \Big (T^2_{\mu \nu} -\frac{T^2}{3} \Big)\Big\vert_{k^2= m^2_2} >0, \;\;\nonumber \\ &&{\mathrm{}} 
\end{eqnarray}
\noindent we come to the conclusion that $$Res(SP)|_{k^2=0} >0, $$ $$ Res(SP)|_{k^2 = m^2_0} >0,$$ $$ Res(SP)|_{k^2= m^2_2} <0,$$ 

\noindent implying that fourth-order gravity is nonunitary  for $D=4$ and {D=5}. 

If $D=3$, the following results are found for the full theory.

\begin{table}[h]
\caption{Signs of the residues of $SP$ at the poles $k^2=0,\;\; k^2= m^2_0, \;\;k^2=m^2_2\;\;$ related to full fourth-order  gravity in three dimensions.} 
\vspace{0.2cm}
\begin{center}
\setlength{\tabcolsep}{10pt} 
{\begin{tabular}{ccc}
\hline
\hline
$D=3$&$\sigma=+1$&$\sigma=-1$\\
\hline
$Res(SP(k))|_{k^2=0}$&$=0$&$=0$\\
$Res(SP(k))|_{k^2=m_{0}^{2}}$&$>0$&$<0$\\
$Res(SP(k))|_{k^2=m_{2}^{2}}$&$<0$&$>0$\\
\hline
\hline
\end{tabular}}
\end{center}
\end{table}

Thus, full tridimensional fourth-order gravity is nonunitaty for $\sigma=\pm 1$. In addition, it is also renormalizable \cite{27}.

NMG, in turn, is tree-level unitary and nonrenormalizable (see Sec. V), while fourth-order gravity in five dimensions with their parameters constrained by the relation  $m^2_0= 9m^2_2$ is nonunitary and nonrenormalizable by power counting.

The preceding results confirm, as expected, that any renormalizable higher-order gravity model is always nonunitary.

\section{Probing the conjecture for $D$-dimensional sixth-order gravity models}
Since we are only interested in the linear part of the action (1), we did not take the $\gamma_0$ term into account. On the other hand, the quadratic part of the resulting action can be written as

\begin{eqnarray}
I^{({\mathrm{sixth-order}})}=\int d^D x \sqrt{|g|} \frac{1}{\kappa^2} \Bigg[ 2R + \frac{1}{2}RF_1(\Box)R + \frac{1}{2}R_{\mu \nu}  F_2(\Box)  R^{\mu \nu}  + \frac{1}{2}R_{\mu \nu \alpha \beta}F_3(\Box) R^{\mu \nu \alpha \beta}\Bigg],
\end{eqnarray}

\noindent where

\begin{eqnarray}
F_1(\Box)\equiv \alpha_0 + \alpha_1 \Box, \; F_2(\Box)\equiv \beta_0 + \beta_1 \Box,\; F_3(\Box)\equiv \gamma_1 \Box. \nonumber
\end{eqnarray}

Now, in the weak field approximation we obtain

\begin{eqnarray} 
R_{\mu \nu \alpha \beta}F_3(\Box)R^{\mu \nu \alpha \beta}= 4R_{\mu \nu}F_3(\Box)R^{\mu \nu} - RF_3(\Box)R  + \partial \Omega + {\cal O}(h^3).
\end{eqnarray}

Substituting (34) into (33) we find

\begin{eqnarray}
I^{({\mathrm{sixth-order}})}= \int d^D x \sqrt{|g|} \frac{1}{\kappa^2} \Bigg[ 2R + \frac{1}{2}R \Big(F_1(\Box) -F_2(\Box) \Big)  R+ \frac{1}{2}R_{\mu \nu} \Big( F_2(\Box) + 4F_3(\Box)\Big) R^{\mu \nu}\Bigg].
\end{eqnarray}

Making the following redefinitions

\begin{eqnarray}
F_1(\Box)- F_3(\Box) \Rightarrow F_1(\Box), \;\;\; F_2(\Box) + 4 F_3(\Box)\Rightarrow F_2(\Box),\nonumber 
\end{eqnarray}

\noindent we come to the conclusion that the quadratic part of our  original  action reduces in this approximation to

\begin{eqnarray}
I^{({\mathrm{sixth-order}})}= \int d^D x \sqrt{|g|} \frac{1}{\kappa^2} \Bigg[ 2R + \frac{\alpha_0 }{2}R^2 + \frac{\beta_0}{2}R^2_{\mu \nu} +  
     \frac{\alpha_1}{2} R \Box R  + \frac{\beta_1}{2} R_{\mu \nu} \Box R^{\mu \nu }
			 \Bigg].
\end{eqnarray}

Taking the same series of actions which we have utilized for verifying our conjecture related to fourth-order gravity models in $D$-dimensions, we find that the propagator concerning sixth-order gravity systems can be written in momentum space as

\begin{eqnarray}
D(k)= &&\Bigg[\frac{1}{k^2} + \frac{1}{m^2_{2_{+}} - m^2_{2_{-}}} \Bigg( \frac{m^2_{2_{-}}}{k^2 - m^2_{2_{+}}} - \frac{m^2_{2_{+}}}{k^2 - m^2_{2_{-}}}  \Bigg) \Bigg]P^{(2)}
 \nonumber \\&&-\frac{1}{D-2}\Bigg[\frac{1}{k^2} + \frac{1}{m^2_{0_{+}} - m^2_{0_{-}}} \Bigg(
\frac{m^2_{0_{-}}}{ k^2 -m^2_{0_{+}}}- \frac{m^2_{0_{+}}}{ k^2 -m^2_{0_{-}}} \Bigg) \Bigg] P^{(0-s)} + (...).
\end{eqnarray}

\noindent Here, $(...)$ stands for the set of terms that are irrelevant for the spectrum of the theory, and

$$m^2_{2_{+}}= \frac{\beta_0}{2 \beta_1} \Bigg(1 \pm \sqrt{1 + \frac{16 \beta_1}{\beta^2_0}} \Bigg),$$
$$ m^2_{0_{\pm}}= \frac{\xi_0}{2\xi_1}\Bigg( 1 \pm \sqrt{1- \frac{4(D-2)\xi_1}{\xi^2_0}} \Bigg), $$

\noindent where $\xi_l = (D-1)\alpha_l + \frac{D}{4} \beta_l  \; (l=0,1).$

As a consequence,

\begin{eqnarray}
{\cal P}_{00,00}({\bf k})= &&- \frac{D-3}{D-2} \frac{1}{{\bf k}^2} + \frac{1}{(D-1)(D-2)} \frac{1}{m^2_{0_{+}} - m^2_{0_{-}}}   \Bigg( \frac{m^2_{0_{-}}}{{ \bf k}^2 + m^2_{0_{+}}} -  \frac{m^2_{0_{+}}}{{\bf k}^2 + m^2_{0_{-}}} \Bigg)\nonumber \\ &&- \frac{D-2}{D-1} 
\frac{1}{m^2_{2_{+}} - m^2_{2_{-}}} \Bigg( \frac{m^2_{2_{-}}}{{\bf k}^2 + m^2_{2_{+}}} - \frac{m^2_{2_{+}}}{{\bf k}^2 + m^2_{2_{-}}} \Bigg). \nonumber
\end{eqnarray}

It follows then that the $D$-dimensional gravitational potential for sixth-order models reads

\begin{eqnarray}
V_3(r)=\frac{\kappa_3 M}{4 \pi}\Bigg\{&& \frac{m^2_{0_{-}}}{m^2_{0_{+}} - m^2_{0_{-}}} K_0(m_{0_{+}}r) - \frac{m^2_{0_{+}}}{m^2_{0_{+}} -m^2_{0_{-}}}  K_0(m_{0_{-}}r) \nonumber\\&&- \frac{m^2_{2_{-}}}{m^2_{2_{+}} - m^2_{2_{-}}}K_0(m_{2_{+}}r) + \frac{m^2_{2_{+}}}{m^2_{2_{+}} - m^2_{2_{-}}} K_0(m_{2_{-}}r)\Bigg\},
\end{eqnarray}

\begin{eqnarray}
V_D(r)=&& -\frac{\kappa_D M}{(2 \pi)^{\frac{D-1}{2}}} \Bigg\{\frac{D-3}{D-2} \frac{2^{\frac{D-5}{2}}}{r^{D-3}} \Gamma \Bigg(\frac{D-3}{2}\Bigg) - \frac{1}{(D-1)(D-2)} \frac{m^2_{0_{-}}}{m^2_{0_{+}} - m^2_{0_{-}}} \Bigg( \frac{m_{0_{+}}}{r }\Bigg)^{\frac{D-3}{2}} \nonumber \\ && \times K_{\frac{D-3}{2}}(m_{0_{+}} r) + \frac{1}{(D-1)(D-2)} \frac{m^2_{0_{+}}}{m^2_{0_{+}} - m^2_{0_{-}}}   \Bigg(\frac{m_{0_{-}}}{r}\Bigg)^{\frac{D-3}{2}} K_{\frac{D-3}{2}}(m_{0_{-}} r) + \frac{D-2}{D-1} \frac{m^2_{2_{-}}}{m^2_{2_{+}} - m^2_{2_{-}}} \nonumber \\ && \times \Bigg(\frac{m_{2_{+}}}{r}\Bigg)^{\frac{D-3}{2}} K_{\frac{D-3}{2}}
(m_{2_{+}}r) - \frac{D-2}{D-1} \frac{m^2_{2_{+}}}{m^2_{2_{+}} - m^2_{2_{-}}}  \Bigg( \frac{m_{2_{-}}}{r}\Bigg)^{\frac{D-3}{2}}K_{\frac{D-3}{2}}(m_{2_{-}} r) \Bigg\}, \;\; (D=4, 5).
\end{eqnarray}

It is trivial to see using (23) that (39) and the Newton gravitational potential coincide for $r \rightarrow \infty$.

Our next step will be to make  a thorough  analysis of the behavior near to the origin of the gravitational potential we have just found.

\subsection{$D=3$}
Taking (24) into account, we find that for $r \ll 1$, (38) assumes the form

\begin{eqnarray}
V_3(r)\sim \frac{\kappa_3 M}{4 \pi} \Bigg(\frac{m^2_{2_{-}} \ln m_{2_{+}} - m^2_{2_{+}} \ln m_{2_{-}}}{m^2_{2_{+}} - m^2_{2_{-}}}  - \frac{m^2_{0_{-}} \ln m_{0_{+}} - m^2_{0_{+}} \ln m_{0_{-}}}{m^2_{0_{+}} - m^2_{0_{-}}} + ... \Bigg).
 \end{eqnarray}

Consequently, the tridimensional sixth-order gravitational potential is finite at the origin and has the following value

\begin{eqnarray}
V_3(0)=\frac{\kappa_3 M}{4 \pi} \Bigg(\frac{m^2_{2_{-}} \ln m_{2_{+}} - m^2_{2_{+}} \ln m_{2_{-}}}{m^2_{2_{+}} - m^2_{2_{-}}}  - \frac{m^2_{0_{-}} \ln m_{0_{+}} - m^2_{0_{+}} \ln m_{0_{-}}}{m^2_{0_{+}} - m^2_{0_{-}}} \Bigg).
 \end{eqnarray}

\subsection{$D=4$}
In this case the gravitational potential is given by

\begin{eqnarray}
V_4(r) = \frac{\kappa_4 M}{4 \pi r}\Bigg(- \frac{1}{2} + \frac{1}{6} \frac{m^2_{0_{-}} 
e^{-m_{0_{+}}r} - m^2_{0_{+}} e^{-m_{0_{-}} r}}{m^2_{0_{+}} -m^2_{0_{-}}} - \frac{2}{3} \frac{m^2_{2_{-}} e^{-m_{2_{+}}r} - m^2_{2_{+}} e^{-m_{2_{-}} r}}{m^2_{2_{+}} -m^2_{2_{-}}} \Bigg).
\end{eqnarray}

Expanding the exponentials at $r=0$, we get

\begin{eqnarray}
V_4(r)\sim\frac{\kappa_4 M}{4 \pi }\Bigg(\frac{2}{3} \frac{m^2_{2_{-}} m_{2_{+}} - m^2_{2_{+}} m_{2_{-}} }{m^2_{2_{+}} -m^2_{2_{-}}} - \frac{1}{6} \frac{m^2_{0_{-}} m_{0_{+}} - m^2_{0_{+}} m_{0_{-}} }{m^2_{0_{+}} -m^2_{0_{-}}}\Bigg) + \; {\cal O}(r).
\end{eqnarray}

Thus, the gravitational potential for sixth-order gravity in four dimensions is finite at the origin, being its value at this point equal to

\begin{eqnarray}
V_4(0)=\frac{\kappa_4 M}{4 \pi }\Bigg(\frac{2}{3} \frac{m^2_{2_{-}} m_{2_{+}} - m^2_{2_{+}} m_{2_{-}} }{m^2_{2_{+}} -m^2_{2_{-}}} - \frac{1}{6} \frac{m^2_{0_{-}} m_{0_{+}} - m^2_{0_{+}} m_{0_{-}} }{m^2_{0_{+}} -m^2_{0_{-}}}\Bigg).
\end{eqnarray}

\subsection{$D=5$}
It is straightforward to show that if $r \ll 1$, (39) reduces, for $D=5$, to

\begin{eqnarray}
V_5(r)\sim \frac{\kappa_5 M}{(2 \pi)^2} \Bigg\{- \frac{3}{8} \frac{m^2_{2_{+}} m^2_{2_{-}}}{m^2_{2_{+}} +m^2_{2_{-}}} \ln \frac{m_{2_{+}}}{m_{2_{-}}}+\frac{1}{24} \frac{m^2_{0_{+}} m^2_{0_{-}}}{m^2_{0_{+}} -m^2_{0_{-}}} \ln \frac{m_{0_{+}}}{m_{0_{-}}}  + ... \Bigg\},
\end{eqnarray}

\noindent which converges to a finite  value at the origin that is equal to

\begin{eqnarray}
V_5(0)= -\frac{\kappa_5 M}{(2 \pi)^2} \Bigg\{ \frac{3}{8} \frac{m^2_{2_{+}} m^2_{2_{-}}}{m^2_{2_{+}} -m^2_{2_{-}}} \ln \frac{m_{2_{+}}}{m_{2_{-}}} - \frac{1}{24} \frac{m^2_{0_{+}} m^2_{0_{-}}}{m^2_{0_{+}} -m^2_{0_{-}}} \ln \frac{m_{0_{+}}}{m_{0_{-}}} \Bigg\}.
\end{eqnarray}

We probe now our conjecture for $D$- dimensional sixth-order gravity models.

\subsection{Verifying the conjecture}
It is not difficult to check by power counting that the superficial divergence related to the system at hand can be written as

\begin{eqnarray}
\delta=D + \frac{6-D}{2}E - \frac{6-D}{2}\sum_{n=3}^{\infty}(n-2)V_n.
\end{eqnarray}

Therefore, we  conclude that

\begin{itemize}
\item $3 \leq D \leq 5 \Rightarrow \delta$ decreases as the number of vertices increase $\Rightarrow$ super-renormalizable 

\item $D=6 \Rightarrow \delta$ is independent of the number of vertices $\Rightarrow$ renormalizable

\item $D \geq 7 \Rightarrow \delta$ increases as the number of vertices increase $\Rightarrow$ nonrenormalizable.

\end{itemize}

Since the gravitational potential can only be computed analytically for $D= 3, 4, 5,$ we restrict our analysis to these dimensions.

On the other hand, we have proved that the gravitational potential  for the full models is finite at $r=0$ in the dimensions above. Accordingly, these models are in total accord with our surmise which requires that they must be non singular at the origin.

For completeness, we finally study the unitarity of the mentioned models.

\subsection{Unitarity of the sixth-order gravity models} 

From (37) we find that the saturated propagator is given by the expression

\begin{eqnarray}
SP(k)=&& \frac{1}{k^2} \Bigg(T_{\mu \nu}T^{\mu \nu} - \frac{1}{D-2}T^2\Bigg) + \Bigg[ \frac{1}{m^2_{2_{+}} -m^2_{2_{-}}} \Bigg( \frac{m^2_{2_{-}}}{k^2 - m^2_{2_{+}}} - \frac{m^2_{2_{+}}}{k^2 - m^2_{2_{-}}} \Bigg) \Bigg]\Bigg(T_{\mu \nu}T^{\mu \nu} - \frac{1}{D-1} T^2 \Bigg) \nonumber \\\ &&-\frac{1}{(D-1)(D-2)} \Bigg[ \frac{1}{m^2_{0_{+}} -m^2_{0_{-}}} \Bigg(  \frac{m^2_{0_{-}}}{k^2 - m^2_{0_{+}}}  - \frac{m^2_{2_{+}}}{k^2 - m^2_{0_{-}}} \Bigg) \Bigg]T^2.
\end{eqnarray} 

So,

\begin{eqnarray}
Res(SP(k))|_{k^2=0}= \Bigg(T_{\mu \nu}T^{\mu \nu} - \frac{1}{D-2}T^2\Bigg)\Bigg|_{k^2=0}, \nonumber
\end{eqnarray}

\begin{eqnarray}
Res(SP(k))|_{k^2=m^2_{2_{+}}}= \frac{m^2_{2_{-}}}{m^2_{2_{+}} - m^2_{2_{-}}}\Bigg(T_{\mu \nu}T^{\mu \nu} - \frac{1}{D-1}T^2\Bigg)\Bigg|_{k^2=m^2_{2_{+}}}, \nonumber
\end{eqnarray}

\begin{eqnarray}
Res(SP(k))|_{k^2=m^2_{2_{-}}}= -\frac{m^2_{2_{+}}}{m^2_{2_{+}} - m^2_{2_{-}}}\Bigg(T_{\mu \nu} T^{\mu \nu} - \frac{1}{D-1}T^2\Bigg)\Bigg|_{k^2=m^2_{2_{-}}}, \nonumber
\end{eqnarray}

\begin{eqnarray}
Res(SP(k))|_{k^2=m^2_{0_{+}}}= - \frac{1}{(D-1)(D-2)} \frac{m^2_{0_{-}}}{m^2_{0_{+}} - m^2_{0_{-}}}  T^2\Big|_{k^2 = m^2_{0_{+}}},
\end{eqnarray}

\begin{eqnarray}
Res(SP(k))|_{k^2=m^2_{0_{-}}}=  \frac{1}{(D-1)(D-2)} \frac{m^2_{0_{+}}}{m^2_{0_{+}} - m^2_{0_{-}}}  T^2\Big|_{k^2 = m^2_{0_{-}}}.
\end{eqnarray}

Our next step is to obtain the signs related to the residues. To do that, however, we need beforehand to know how $m^2_{2_{+}}$ and $m^2_{2_{-}}$, as well as  $m^2_{0_{+}}$ and $m^2_{0_{-}}$ are ordered. To facilitate this task, we redefine the following parameters

\begin{eqnarray}
\alpha_0 \mapsto \kappa^2 \alpha_0, \; \alpha_1 \mapsto \kappa^4 \alpha_1, \; \beta_0 \mapsto \kappa^2 \beta_0, \;\beta_1 \mapsto \kappa^4 \beta_1. \nonumber 
\end{eqnarray}
\noindent which implies that in terms of these redefined parameters the masses $m^2_{_{\pm}}$ and $m^2_{0_{\pm}}$ assume the form

\begin{align} \label{massas}
m^2_{2_{\pm}} = \frac{\beta_0}{2 \kappa^2  \beta_1} \left( 1 \pm \sqrt{1 + \frac{16\beta_1}{\beta_0^2}} \right), \qquad
\end{align}

\begin{align}
m^2_{0_{\pm}} = \frac{\xi_0}{2\kappa^2  \xi_1} \left( 1 \pm  \sqrt{1 -  \frac{4(D-2)\xi_1}{\xi_0^2} } \right),
\end{align}
where $\xi_l = 3\alpha_l + \beta_l$ ($l=0,1$). Actually, we are interestd in the following regions in the  parametric spaces  

\begin{eqnarray}
\Omega_{\beta} = \Bigg\{(\beta_0,\beta_1) \in \mathbb{R}^2 \,\Big| \,\, \kappa^2 m^2_{2_{+}} > 0 \,\,\, \textmd{and}\,\,\, \kappa^2 m^2_{2_{-}} > 0 \Bigg\}, \nonumber
\end{eqnarray}
\begin{eqnarray}
\Omega_{\xi} = \Bigg\{(\xi_0,\xi_1) \in \mathbb{R}^2 \,\Big| \,\, \kappa^2 m^2_{0_{+}} > 0 \,\,\, \textmd{and}\,\,\, \kappa^2 m^2_{0_{-}} > 0 \Bigg\}, \nonumber
\end{eqnarray}

\begin{eqnarray}
\Omega_{\alpha} = \Bigg\{ (\alpha_0,\alpha_1) = \Bigg(\frac{4 \xi_0 - D\beta_0}{4(D-1)},\frac{4\xi_1 - D \beta_1}{4(D-1)} \Bigg) \in \mathbb{R}^2  \Big| \,\, (\beta_0,\beta_1) \in \Omega_\beta \,\,\, {\mathrm {and}} \,\,\, (\xi_0,\xi_1) \in \Omega_\xi \Bigg\}. \nonumber
\end{eqnarray} 
 Taking (51) and (52) into account we may write

\begin{eqnarray}
\Omega_{\beta} = \Bigg\{(\beta_0,\beta_1) \in \mathbb{R}^2 \, \Big| \,\, \beta_0 < 0 \,\,\,\, \textmd{and}\,\, -\beta_0^2/16 < \beta_1 <0 \Bigg \}, \nonumber
\end{eqnarray}
\begin{eqnarray}
\Omega_{\xi} = \Bigg\{(\xi_0,\xi_1) \in \mathbb{R}^2 \,\Big| \,\, \xi_0 > 0 \,\,\, \textmd{and}\,\,\, 0 < \xi_1 < \xi_0^2/4(D-2) \Bigg\}.\nonumber
\end{eqnarray}

As a result, we find that in these regions the masses are ordered as 
\begin{eqnarray}\label{ordenamento}
m^2_{2_{+}} > m^2_{2_{-}} \;\; {\mathrm{and}} \;\; m^2_{0_{+}} > m^2_{0_{-}}.
\end{eqnarray}

Now, from (32) and (53), we arrive at the conclusion that

\begin{subequations}
\begin{eqnarray}
Res\left( SP(k) \right)|_{k^2 =0} > 0, \nonumber
\end{eqnarray}
\begin{eqnarray}
Res\left( SP(k) \right)|_{k^2 =m^2_{2_{+}}} >0, \qquad Res\left( SP(k) \right)|_{k^2 =m^2_{2_{-}}} < 0, \nonumber
\end{eqnarray}
\begin{eqnarray}
Res\left( SP(k) \right)|_{k^2 =m^2_{0_{+}}} <0, \qquad Res\left( SP(k) \right)|_{k^2 =m^2_{0_{-}}} > 0. \nonumber
\end{eqnarray}
\end{subequations}

Consequently, the particle content of the model is made up of three healthy particles and two ghosts, which clearly shows that  full sixth-order gravity is nonunitary.

The results above confirm once more that renormalizable higher-order gravity models are nonunitarry.

\section{Final comments}
We have verified that renormalizable higher-order gravitational models, specifically fourth- and sixth- order gravity systems in $D$- dimensions, possess a singularity free classical potential at the origin. The converse  is not necessarily true. Indeed, consider the  gravity system  in four dimensions defined by the Lagrangian \cite{16}

\begin{eqnarray}
{\cal L}= \sqrt{|g|}\Big( \frac{2}{\kappa^2} R + \alpha_0 R^2  + a_1 R \Box R + b_0 R^2_{\mu \nu}
\Big), \nonumber
\end{eqnarray}

\noindent wherein the masses of the modes related to higher-order terms are given by

\begin{eqnarray}
m^2_{(0)\pm}= \frac{3a_0 + b_0 \pm \sqrt{(3a_0 + b_0)^2 -24a_1 \kappa^{-2}}}{6 a_1}, \qquad\qquad m^2_{(2)} = \frac{4}{|b_0|\kappa^2}.
\end{eqnarray}

\noindent Here $m_{(2)}$ and $m_{(0)+}$ are ghost excitations, while $m_{(0)-}$ is a healthy mode \cite{12}.

In this scenario the potential is given by
\begin{eqnarray}
V_4(r)= - \frac{G_4 M}{r} \Bigg[ 1 - \frac{4}{3} e^{- m_{(2)} r} + \frac{1}{3}\Bigg(\frac{m^2_{(0)-}}{m^2_{(0)-} - m^2_{(0)+}}  e^{- m_{(0)+}r} + \frac{m^2_{(0)+}}{m^2_{(0)+} - m^2_{(0)-}} e^{- m_{(0)-}r} \Bigg) \Bigg],
\end{eqnarray} 

\noindent and, as a consequence, in the region near the origin it assumes the form

\begin{eqnarray}
V_4(r)\sim G_4 M \Bigg[- \frac{4}{3}m_{(2)} + \frac{1}{3} \frac{m_{(0)+} m_{(0)-} - (m_{(0)+} -m_{(0)-})}{m^2_{(0)+} - m^2_{(0)-}} \Bigg] + {\cal O}(r).
\end{eqnarray}

Therefore, the potential is finite at $r=0$. Nonetheless, the model at hand is nonrenormalizable by power counting which implies that the finiteness of the classical potential at the origin  is a necessary, but certainly not a sufficient condition for the renormalizability of the model.

 In summary,  if a higher-derivative  gravity model is renormalizable,  it is necessarily  nonunitary and, in addition, is endowed with a classical potential finite at the origin, but the opposite is not true in general.

We have also confirmed the general premise that renormalizable higher-derivativ gravity models are nonunitary.

Now, we address ourselves to the issue of NMG \cite{22}. Our main interest in this system is owed to the fact that it was  by  analyzing its properties that the idea of the conjecture came to light.  As is well known, this model aroused  a great interest in the physical community when it was conceived since it is a tree-level unitary higher-order gravity model; in fact, tree-level unitary higher-derivative gravity systems are extremely rare in physics. On the other hand, the aforementioned  theory  caused considerable controversy as far as its renormalizability is concerned. Really, it was initially claimed to be renormalizable by Oda \cite{26}, being some years late shown to be nonrenormalizable by Muneyuki and Ohta \cite{27}.
It is exactly the disagreement between these results that we want to discuss in the framework of our conjecture. Nevertheless, for clarity's sake,  we begin by presenting some important points related to to the system at hand.

\subsection{Tree-level unitarity}
From (17) it is straightforward to obtain the saturated propagator, i.e.

\begin{eqnarray}
SP(k)= \frac{1}{\sigma}\Big[\frac{1}{k^2} - \frac{1}{k^2- m^2_2}\Big] \Big[T^2_{\mu \nu} - \frac{1}{2}T^2\Big]  + \frac{1}{\sigma}\Big[- \frac{1}{k^2} + \frac{1}{k^2 - m^2_0}\Big] \frac{1}{2}T^2.
\end{eqnarray}

Eq. (18), in trurn, furnishes the constraints

\begin{eqnarray}
\frac{\sigma}{\beta}<0, \;\;\;\; \frac{\sigma}{8\alpha + 3 \beta}>0.
\end{eqnarray}

Now, the residues of $SP(k)$ at the poles $k^2=m^2_2,\; k^2=0,$ and $k^2= m^2_0$ are, respectively,

\begin{eqnarray}
Res(SP)|_{k^2= m^2_2}= - \frac{1}{\sigma}\Big(T^2_{\mu \nu} - \frac{1}{2}T^2 \Big)\Big|_{k^2= m^2_2},
\end{eqnarray}

\begin{eqnarray}
Res(SP)|_{k^2= 0}= - \frac{1}{\sigma}\Big(T^2_{\mu \nu} - T^2 \Big)\Big|_{k^2= 0},
\end{eqnarray}

\begin{eqnarray}
Res(SP)|_{k^2= m^2_0}= - \frac{1}{2\sigma}\Big( T^2 \Big)\Big|_{k^2= m^2_0}.
\end{eqnarray}
Thence, we arrive at the conclusion that (i) $Res(SP)|_{k^2 = m^2_2} >0$ if $\sigma=-1$ (which we assume to be  the case from now on), and $Res(SP)|_{k^2=0}$. As a result, we need not worry about these poles, the troublesome one is $k^2= m^2_0$ since $Res(SP)|_{k^2= m^2_0} <0$. A way out of this difficult is to consider the $m_0 \rightarrow \infty$ limit of the model under discussion, which leads us to conclude that $\alpha= -\frac{3}{8}\beta$. Accordingly, the class of models defined by the Lagrangian,

\begin{eqnarray}
{\cal L}= \sqrt{|g|}\Bigg[ -\frac{2R}{\kappa^2} + \frac{\beta}{2} \Bigg(R^2_{\mu \nu} - \frac{3}{8}R^2\Bigg) \Bigg],
\end{eqnarray}

\noindent where $\kappa^2= 4 \kappa_3$, are ghost free at the tree level. For convenience's sake, we replace $\beta$ with $\frac{4}{\kappa^2 m^2_2}$. The resulting Lagrangian,

\begin{eqnarray}
{\cal L}_{\mathrm{NMG}}= \sqrt{|g|}\Bigg[ -\frac{2R}{\kappa^2} + \frac{2}{\kappa^2 m^2_2} \Bigg(R^2_{\mu \nu} - \frac{3}{8}R^2\Bigg) \Bigg],
\end{eqnarray} 

\noindent defines the famous system baptized  New Massive Gravity \cite{22, 23, 24, 25}.

At this point it is interesting to recall some comments that in a sense predicted the nonrenormalizability of NMG.

\begin{itemize}
\item It is not clear at all whether or not the particular ratio between $\alpha$ and 
$\beta$ will survive renormalization at a given loop, even at one loop; in other words, unitarity beyond the tree level has to be checked \cite{36}.

\item Most likely, NMG is nonrenormalizable since it only improves the spin-2 projections of the propagator but not the spin-0 projection \cite{37}.
 \end{itemize}
Undoubtedly, these remarks anticipated for a few years the definitive proof related to the nonrenormalizability of NMG.

\subsection{Gravitational potential}

From (22) we get without any difficult

\begin{eqnarray}
V_{\mathrm{NMG}}(r)= -\frac{\kappa_3 M}{4 \pi}K_0(m_2 r).
\end{eqnarray}

Note that the potential concerning NMG has a logarithm  singularity at the origin.

\subsection{Discussing the renormalizability of NMG via our conjecture} 
According to Oda \cite{26}, NMG is renormalizable. Nevertheless, this author made a  mistake when he considered NMG as a full three dimensional gravity model (with $\sigma =-1$), being the latter  renormalizable. In other words, although the birth of  NMG is the full gravity model just mentioned (see Fig. 1), the system under discussion has  a constraint between its parameters ($\alpha= - \frac{3}{8}\beta$). It is exactly this special relation between the parameters the responsible for breaking the renormalizability of the full model as it was demonstrated by Muneyuki and Ohta \cite{27}.

Examining the diagram depicted in Fig. 1, we clearly see that as $m_0$ becomes greater and greater, the full potential $V_3(r)$ with $\sigma=-1$ and $m_2 < m_0$ (see (22)) rapidly approaches  the potential concerning  NMG and eventually  they coalesce. It worth mentioning  that to arrive at  the NMG potential from the full potential above, the latter must necessarily become singular at the origin which takes place in the $m_0 \rightarrow \infty$ limit. It is remarkable that  this is precisely the condition for avoiding  at the tree level, the massive spin-0 ghost that haunts full tridimensional fourth-order gravity. Accordingly, the presence of the singularity in NMG is correlated to the absence of the tree-level ghost; which means that the renormalizability of the model and its consequent nonunitary,  and the existence of a singularity in the potential are  intertwined. In the diagram shown in Fig. II, the behavior of full fourth-order gravity in three dimension is depicted as far as its unitarity,  renormalizability, and the existence of a finite gravitational potential at the origin, are concerned. A cursory glance at this diagram  suggests that in three dimensions a unitary system is nonrernormalizable, being  connected to a singular potential at  the origin, while a renormalizable model is related to to a potential finite at the origin,  being in addition nonunitary. Interestingly enough, it was exactly  the analysis of this model that led us to propose the conjecture analyzed in this paper.

\begin{figure}[htb!]
	\centering
		\includegraphics[scale=0.8]{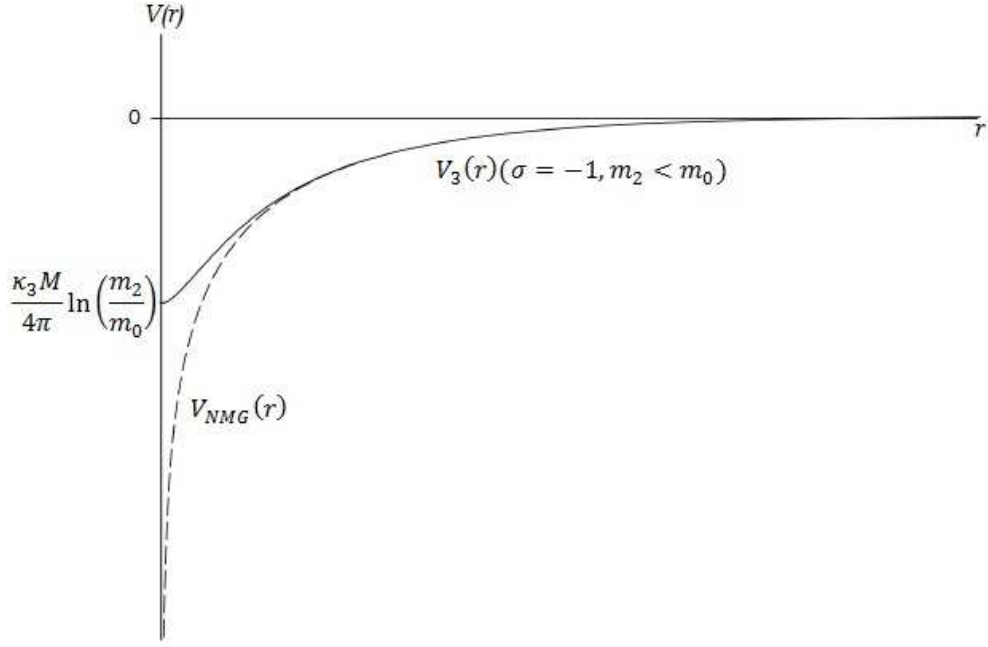}
	\caption{Gravitational potential for both  the full fourth-order gravity model  in three dimensions with $\sigma=-1$ and $m_2 < m_0$ (continuous line) and NMG (dashed line).}
	\label{fig}
\end{figure}

 \begin{figure}[htb!]
	\centering
		\includegraphics[scale=0.8]{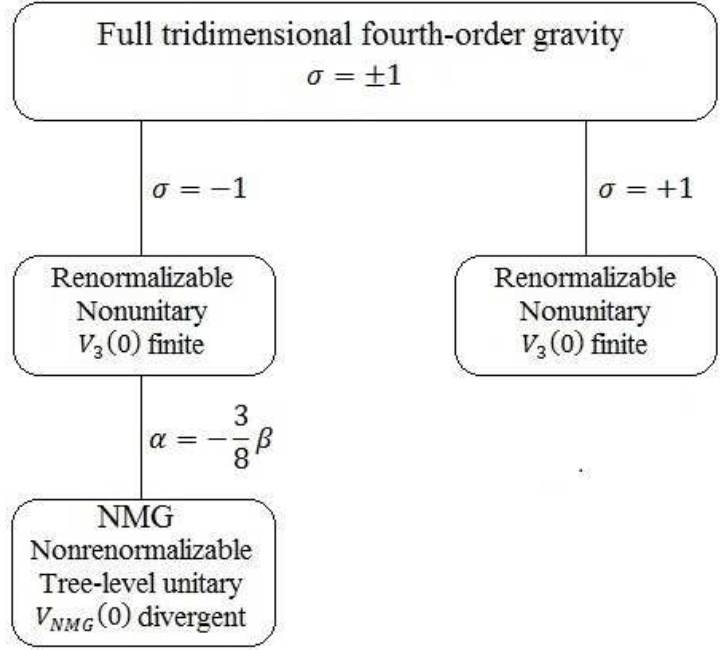}
	\caption{Renormalizability, unitarity, and the gravitational potential at the origin concerning full fourth-order gravity in three dimensions ($\sigma =\pm1$) .}
	\label{fig}
\end{figure}

Last but not least we remark that although we have only tested our premise  for some particular $D$-dimensional higher-derivative gravitational models, the surmise is completely general. In fact, our conjecture is valid for the most general $D$-dimensional gravitational action below

\begin{eqnarray}
I_D= \int d^D x \sqrt{|g|}\Bigg( \frac{2 \sigma}{\kappa^2}R + \frac{1}{2 \kappa^2} RF_1(\Box)R    +\frac{1}{2\kappa^2}R^{\mu \nu} F_2(\Box)R_{\mu\nu} + \frac{1}{2}R_{\mu \nu \alpha \beta}F_3(\Box)R^{\mu \nu \alpha \beta} \Bigg).\nonumber
\end{eqnarray}
\noindent Here,

\begin{eqnarray}
F_1(\Box)=\sum_{n=0}^p \alpha_n (\Box)^n + f_1(\Box),
\end{eqnarray}

\begin{eqnarray}
F_2(\Box)=\sum_{n=0}^q \beta_n (\Box)^n + f_2(\Box),
\end{eqnarray}

\begin{eqnarray}
F_3(\Box)=\sum_{n=0}^r \gamma_n (\Box)^n + f_3(\Box).
\end{eqnarray}

\noindent where $f_1(\Box) $,  $f_2(\Box) $, and  $f_3(\Box)$ are nonlocal functions, and  $\alpha_n \; (n=0, ...,p)$,  $\beta_n \; (n=0, ...,q)$  and $\gamma_n \; (n=0, ...,r)$ are real coefficients. These results will be published elsewhere \cite{38}.

\begin{acknowledgments}
The authors are very grateful to CNPq  and  FAPERJ,  for financial support.
\end{acknowledgments}

\appendix
\section{{D}-DIMENSIONAL EINSTEIN CONSTANT}
As is well known, the $D$-dimensional Poisson equation can be written as

\begin{eqnarray}
\nabla^2_{D-1} \varphi_D({\bf x})= G_D \frac{2 \pi^{\frac{D-1}{2}}}{\Gamma\Big (\frac{D-1}{2}\Big)}\rho,
\end{eqnarray}
\noindent where $\rho$ is the mass density.

On the other hand, the Schwarzschild metric in isotropic coordinates reads

\begin{eqnarray}
 ds^2=\Bigg[\frac{1 + \frac{1}{2}\varphi_D({ \bf x})}{ 1 - \frac{1}{2} \varphi_D({\bf x})}\Bigg]^2 dt^2 - \Bigg[ 1 - \frac{1}{2} \varphi_D({\bf x})\Bigg]^{\frac{4}{D-3}}\Bigg[ \Big(dx^1\Big)^2 + ... + \Big(d x^{D-1}\Big)^2\Bigg].
\end{eqnarray}

In the Newtonian limit, i.e. far form the mass distributions, the previous metric assumes the form

\begin{eqnarray} 
ds^2= \Big[ 1 +2 \varphi_D({\bf x })\Big]dt^2 -\Bigg[1- \frac{2}{D-3} \varphi_D{({\bf x})} \Bigg]  \Bigg[ \Big(dx^1\Big)^2 + ... + \Big(d x^{D-1}\Big)^2\Bigg].
\end{eqnarray}

From the  Einstein equations, namely $G_{\mu \nu
}= \kappa_D T_{\mu\nu}$, we then find

\begin{eqnarray}
G_{00}= \kappa_D \rho= \frac{D-2}{D-3}\nabla^2_{D-1 }\varphi_{D}(\bf x).
\end{eqnarray}

Therefore, we come to the conclusion that

\begin{eqnarray}
\kappa_D= \frac{D-2}{D-3} G_D  \frac{2 \pi^{ \frac{D-1}{2}}}{\Gamma\Big( \frac{D-1}{2}\Big)} \;\;\;(D>3).
\end{eqnarray} 

As we have  already commented in the Introduction, in $D=3$, $\kappa_3$ cannot be related to $G_3$; nonetheless, for simplicity's sake $\kappa_3$ is used in general as the symbol for the  tridimensional Einstein constant, although it is unrelated to $G_3$.

\section{$D$-DIMENSIONAL BARNES-RIVERS OPERATORS }
The complete set of the $D$-dimensional Barnes-Rivers operators in momentum space is given by
\begin{eqnarray}
P^{(2)}_{\mu\nu, \kappa \lambda}= \frac{1}{2}\Big(\theta_{\mu \kappa} \theta_{\nu \lambda} + \theta_{\mu \lambda} \theta_{\nu \kappa}\Big) - \frac{1}{D-1} \theta_{\mu \nu}\theta_{\kappa \lambda }, \nonumber
\end{eqnarray}

\begin{eqnarray}
P^{(1)}_{\mu\nu, \kappa \lambda}= \frac{1}{2}\Big(\theta_{\mu \kappa} \omega_{\nu \lambda} + \theta_{\mu \lambda} \omega_{\nu \kappa} + \theta_{\nu \lambda} \omega_{\mu\kappa} + \theta_{\nu \kappa}\omega_{ \mu \lambda } \Big), \nonumber
\end{eqnarray}

\begin{eqnarray}
P^{(0-s)}_{\mu \nu, \kappa \lambda}= \frac{1}{D-1}\theta_{\mu\nu} \theta_{
\kappa \lambda}, \;\;\; 
P^{(0-w)}_{\mu \nu, \kappa \lambda}= \frac{1}{D-1}\omega_{\mu \nu} \omega_{
\kappa \lambda}, \nonumber
\end{eqnarray}

\begin{eqnarray}
P^{(0-sw)}_{\mu \nu, \kappa \lambda}= \frac{1}{\sqrt{D-1}}\theta_{\mu\nu} \omega_{
\kappa \lambda}, \;\;\; 
P^{(0-ws)}_{\mu \nu, \kappa \lambda}= \frac{1}{\sqrt{D-1}}\omega_{\mu \nu} \theta_{
\kappa \lambda}, \nonumber
\end{eqnarray}
\noindent where $\theta_{\mu \nu}\equiv \eta_{\mu \nu} -\frac{k_\mu k_\nu}{k^2}$ and $
\omega_{\mu \nu} \equiv \frac{k_\mu k_\nu}{k^2}$ are, respectively, the usual transverse and longitudinal vectorial  projection operators. The multiplicative table for these operators is displayed in Table II. 

\begin{table}
\caption{Multiplicative table for the Barnes-Rivers operators.} 
\vspace{0.2cm}
\centering
\setlength{\tabcolsep}{8pt} 
{\begin{tabular}{@{}*{7}{l}}
\hline
\hline
&$P^{(2)}$&$P^{(1)}$&$P^{(0-s)}$&$P^{(0-w)}$&$P^{(0-sw)}$&$P^{(0-ws)}$\\
\hline
$P^{(2)}$&$P^{(2)}$&0&0&0&0&0\\
$P^{(1)}$&0&$P^{(1)}$&0&0&0&0\\
	$P^{(0-s)}$&0&0&$P^{(0-s)}$&0&$P^{(0-sw)}$&0\\
$P^{(0-w)}$&0&0&0&$P^{(0-w)}$&0&$P^{(0-ws)}$\\
$P^{(0-sw)}$&0&0&0&$P^{(0-sw)}$&0&$P^{(0-s)}$\\
$P^{(0-ws)}$&0&0&$P^{(0-ws)}$&0&$P^{(0-w)}$&0\\
\hline
\hline
\end{tabular}}

\end{table}

\section{SOME RELEVANT INTEGRALS}
The integrals related to the models dealt with in the article can be generically written as  
\begin{eqnarray}
\int \frac{d^{D-1}{\bf k}}{(2 \pi)^{D-1}}f(|{\bf k|}) e^{i {\bf k} \cdot {\bf r}}.
\end{eqnarray}
Now, keeping in mind that
\begin{eqnarray}
\int \frac{d^{D-1}{\bf k}}{(2 \pi)^{D-1}}f(|{\bf k|}) e^{i {\bf k} \cdot {\bf r}}= \frac{1}{(2 \pi)^{\frac{D-1}{2}}} \frac{1}{r^{\frac{D-3}{2}}}  \int_0^{\infty}x^{\frac{D-1}{2}}f(x)J_{\frac{D-3}{2}}(xr) dx  \qquad (D>2), \;\; ({\mathrm{see}} \;\;{\mathrm{ Ref}}. \;13) \nonumber
\end{eqnarray}

\noindent where $x\equiv|{\bf k}|$, we promptly find the following results

\begin{eqnarray}
 \int \frac{d^{D-1}{\bf k}}{(2 \pi)^{D-1}}\frac{ e^{i {\bf k} \cdot {\bf r}}}{{\bf k}^2}=\frac{1}{(2\pi)^{\frac{D-1}{2}}} \frac{1}{r^{D-3}} \int_0
^{\infty} y^{\frac{D-5}{2}} J_{\frac{D-3}{2}}(y)dy = \frac{1}{(2 \pi)^{\frac{D-1}{2}}} \frac{1}{r^{D-3}}I_D, \nonumber
\end{eqnarray}

\begin{eqnarray}
 \int \frac{d^{D-1}{\bf k}}{(2 \pi)^{D-1}}\frac{ e^{i {\bf k} \cdot {\bf r}}}{{\bf k}^2 + m^2}=\frac{1}{(2\pi)^{\frac{D-1}{2}}} \frac{1}{r^{D-3}} \int_0
^{\infty}\frac{ y^{\frac{D-1}{2}}}{y^2 + m^2 r^2}  J_{\frac{D-3}{2}}(y)dy = \frac{1}{(2 \pi)^{\frac{D-1}{2}}} \frac{1}{r^{D-3}}{\cal I}_D(r). \nonumber 
\end{eqnarray}  

\noindent Here,

\begin{eqnarray}
I_D\equiv \int_0^{\infty} y^{\frac{D-5}{2}}J_{\frac{D-3}{2}}(y) dy, 
\end{eqnarray} 
 \noindent and  
\begin{eqnarray}
{\cal I}(r) \equiv &&\int_0^{\infty} \frac{y^{\frac{D-1}{2}}}{y^2 + m^2r^2}  J_{\frac{D-3}{2}}(y)dy. \nonumber
\end{eqnarray}

From the Table of integrals, Series, and Products  by Gradshteyn and Ryzhik \cite {39}, we obtain

\begin{eqnarray}
I_D= 2^{\frac{D-5}{2}}\Gamma \Bigg(\frac{D-3}{2}\Bigg), \;\;\; (D=4,5) 
\end{eqnarray}

\begin{eqnarray}
{\cal I}_D(r)= (mr)^{\frac{D-3}{2}}K_{\frac{D-3}{2}}(mr).\;\;\; (D=3,4,5) 
\end{eqnarray}

Accordingly,

\begin{eqnarray}
\int \frac{d^{D-1}{\bf k}}{(2\pi)^{D-1}} \frac{e^{i {\bf k}\cdot {\bf r}}}{{{\bf k}}^2}= \frac{1}{(2 \pi)^{
\frac{D-1}{2}}} \frac{2^{\frac{D-5}{2}}}{r^{D-3}} \Gamma \Bigg( \frac{D-3}{2}\Bigg), \;\;\; (D=4,5) \nonumber
\end{eqnarray}

\begin{eqnarray}
\int \frac{d^{D-1}{\bf k}}{(2\pi)^{D-1}} \frac{e^{i {\bf k}\cdot {\bf r}}}{{{\bf k}}^2 + m^2}=&& \frac{1}{(2 \pi)^{
\frac{D-1}{2}}} \Bigg(\frac{m}{r}\Bigg)^{\frac{D-3}{2}} K_{\frac{D-3}{2}}(mr).  \nonumber \\ && (D=3,4,5) \nonumber
\end{eqnarray}

\end{document}